\documentclass[prd,nofootinbib]{revtex4}%
\usepackage{amsfonts}
\usepackage{latexsym}
\usepackage{epsfig}
\usepackage{amssymb}
\usepackage{amsmath}
\usepackage{graphicx}%
\begin{document}
\title{Electric Charges and Magnetic Monopoles in Gravity's Rainbow}
\author{Remo Garattini}
\email{Remo.Garattini@unibg.it}
\affiliation{Universit\`{a} degli Studi di Bergamo, Dipartimento di Ingegneria,}
\affiliation{Viale Marconi 5, 24044 Dalmine (Bergamo) Italy}
\affiliation{I.N.F.N. - sezione di Milano, Milan, Italy.}
\author{Barun Majumder}
\email{barunbasanta@iitgn.ac.in}
\affiliation{Indian Institute of Technology Gandhinagar}
\affiliation{Ahmedabad, Gujarat 382424, India}

\begin{abstract}
In this work, we explore the possibility that quantum fluctuations induce an
electric or magnetic charge or both, in the context of Gravity's Rainbow. A
semi-classical approach is adopted, where the graviton one-loop contribution
to a classical energy in a background spacetime is computed through a
variational approach with Gaussian trial wave functionals. The energy density
of the graviton one-loop contribution, in this context, acts as a source for
the electric/magnetic charge. The ultraviolet (UV) divergences, which arise
analyzing this procedure, are kept under control with the help of an
appropriate choice of the Rainbow's functions. In this way we avoid the
introduction of any regularization/renormalization scheme. A comparison with
the observed data leads us to determine the size of the electron and of the
magnetic monopole which appear to be of Planckian size. Both results seem to
be of the same order for a Schwarzschild and a de Sitter background,
respectively. Estimates on the magnetic monopole size have been done with the
help of the Dirac quantization procedure. We find that the monopole radius is
larger than the electron radius. Even in this case the ratio between the
electric and magnetic monopole radius appears to be of the same order for both geometries.

\end{abstract}
\maketitle

\section{Introduction}

It was Andrei Sakharov in 1967\cite{Sakharov} who first conjectured the idea
of \textit{Induced gravity} (or emergent gravity), namely space-time
background emerges as a mean field approximation of underlying microscopic
degrees of freedom, similar to the fluid mechanics approximation of
Bose--Einstein condensates. This means that some basic ingredients of General
Relativity like the gravitational Newton's constant can be computed by means
of quantum fluctuations of some matter fields. This idea is opposed to the
concept of \textquotedblleft\textit{charge without charge}\textquotedblright%
\ and \textquotedblleft\textit{mass without mass}\textquotedblright\ arising
from the \textit{spacetime foam} picture of John A. Wheeler\cite{Wheeler},
where the matter properties emerge as a geometrical feature of space time. In
a \textit{foamy spacetime} topological fluctuation appear at the Planck scale,
meaning that spacetime itself undergoes a deep and rapid transformation in its
structure. Wheeler also considered wormhole-type solutions as objects of the
quantum spacetime foam connecting different regions of spacetime at the Planck
scale. Although the Sakharov approach has the appealing property of being
renormalizable \textquotedblleft\textit{ab initio}\textquotedblright\ because
it involves only quantum fluctuations of matter fields described by bosons and
fermions, the Wheeler picture involves quantum fluctuations of the
gravitational field alone and since one of the purposes of Quantum Gravity
should be a realization of a theory combining Quantum Field Theory with
General Relativity, it appears that spacetime foam is the right candidate for
such a description. Unfortunately, every proposal of Quantum Gravity except
string theory has to face with Ultra Violet (UV) divergences. Recently a
proposal which uses a distortion of the gravitational field at the Planck
scale, named as \textit{Gravity's Rainbow}\cite{GAC,MagSmo} has been
considered to compute Zero Point Energy (ZPE) to one loop\cite{GaMa,Remof(R)}.
The interesting point is that such a distortion enters into the background
metric and becomes active at Planck's scale keeping under control UV
divergences. Briefly, the situation is the following: one introduces two
arbitrary functions $g_{1}\left(  E/E_{P}\right)  $ and $g_{2}\left(
E/E_{P}\right)  $, denoted as \textit{Rainbow's functions}, with the only
assumption that
\begin{equation}
\lim_{E/E_{P}\rightarrow0}g_{1}\left(  E/E_{P}\right)  =1\qquad\text{and}%
\qquad\lim_{E/E_{P}\rightarrow0}g_{2}\left(  E/E_{P}\right)  =1.\label{lim}%
\end{equation}
On a general spherical symmetric metric such functions come into play in the
following manner
\begin{equation}
ds^{2}=-N^{2}\left(  r\right)  \frac{dt^{2}}{g_{1}^{2}\left(  E/E_{P}\right)
}+\frac{dr^{2}}{g_{2}^{2}\left(  E/E_{P}\right)  \left(  1-\frac{b\left(
r\right)  }{r}\right)  }+\frac{r^{2}}{g_{2}^{2}\left(  E/E_{P}\right)
}\left(  d\theta^{2}+\sin^{2}\theta d\phi^{2}\right)  \,,\label{dS}%
\end{equation}
where $N(r)$ is the lapse function, $b(r)$ is denoted as the shape function
and $E_{P}$ is the Planck energy. The purpose of this paper is to approach one
of the aspects of Wheeler's ideas, namely \textquotedblleft\textit{charge
without charge}\textquotedblright. In particular, we will investigate if
quantum fluctuations of the gravitational field can be considered as a source
for the electric/magnetic charge. Note that a similar approach to realize
\textquotedblleft\textit{charge without charge}\textquotedblright\ has been
described in Ref.\cite{RemoCharge}. However due to UV divergences a
regularization/renormalization was used to obtain finite results and if a
renormalization group like equation has been used, the final result would
depend on the renormalization point scale $\mu_{0}$. Here the renormalization
point is fixed at the Planck scale due the Rainbow's functions. It is clear
that, if an electric/magnetic charge can be generated, this information is
encoded in the Einstein's field equations. These equations are simply
summarized by
\begin{equation}
R_{\mu\nu}-\frac{1}{2}g_{\mu\nu}R=\kappa T_{\mu\nu},\label{ein}%
\end{equation}
where%
\begin{equation}
T_{\mu\nu}=\frac{1}{4\pi}\left[  F_{\mu\gamma}F_{\nu}^{\gamma}-\frac{1}%
{4}g_{\mu\nu}F_{\gamma\delta}F^{\gamma\delta}\right]  \label{EM}%
\end{equation}
is the energy-momentum tensor of the electromagnetic field, $F_{\mu\nu
}=\partial_{\mu}A_{\nu}-\partial_{\nu}A_{\mu}$, $\kappa=8\pi G$ with $G$ the
Newton's constant and here we have neglected the contribution of the
cosmological constant $\Lambda_{c}$. The electromagnetic field strength tensor
$F_{\mu\nu}$ can be computed with the help of the electromagnetic potential
$A_{\mu}$ which, in the case of a pure electric field assumes the form
$A_{\mu}=\left(  Q_{e}/r,0,0,0\right)  $ while in the case of pure magnetic
field, the form is $A_{\mu}=\left(  0,0,0,-Q_{m}\cos\theta\right)  $. $Q_{e}$
and $Q_{m}$ are the electric and magnetic charge respectively. It is
interesting to note that $Q_{e}$ and $Q_{m}$ contribute in the same way to the
electromagnetic Hamiltonian density. Indeed, for the electric charge, the
on-shell contribution of $T_{\alpha\beta}u^{\alpha}u^{\beta}$ is
\begin{equation}
T_{\mu\nu}u^{\mu}u^{\nu}=\frac{1}{8\pi}\left(  F_{01}\right)  ^{2}=\frac
{1}{8\pi}\frac{Q_{e}^{2}}{r^{4}}=\rho_{e},\label{elec}%
\end{equation}
and when we consider the magnetic charge, we get
\begin{equation}
T_{\mu\nu}u^{\mu}u^{\nu}=\frac{1}{8\pi}\left(  F_{23}\right)  ^{2}=\frac
{1}{8\pi}\frac{Q_{m}^{2}}{r^{4}}=\rho_{m}.\label{magn}%
\end{equation}
$u^{\mu}$ is a time-like unit vector such that $u\cdot u=-1$. However, while
the electric charge exists, for the magnetic charge or magnetic monopole,
there is no experimental evidence for its existence\footnote{Recently, it has
been discovered that spin ices, frustrated magnetic systems, have effective
quasiparticle excitations with magnetic charges very close to magnetic
monopoles\cite{SpinIce}.}. The magnetic monopole search has a long history in
theoretical physics: predicted by Paul Dirac in 1931, he showed that QED
allows the existence of point-like magnetic monopole with charge%
\begin{equation}
Q_{m}=\frac{2\pi}{Q_{e}}\label{ECMM}%
\end{equation}
or an integer multiple of it\cite{m1}. Subsequently this prediction was also
confirmed by Gerard 't Hooft and Alexander Polyakov who showed that magnetic
monopoles are predicted by all Grand Unified Theories (GUTs) \cite{m2}.
Although monopoles of grand unified theories would have masses typically of
the order of the unification scale ($m\sim10^{16}$ GeV) but generally there
are no tight theoretical constraints on the mass of a monopole. For this
reason, the reference value of our calculation will be that of the electric
charge. It is important to remark that in a system of units in which
$\hbar=c=k=1$, that will be used throughout the paper\footnote{For example, in
SI units%
\begin{equation}
\frac{e^{2}}{4\pi\hbar c\epsilon_{0}}=\frac{1}{137}.
\end{equation}
}%
\begin{equation}
e^{2}=\frac{1}{137}.
\end{equation}
The rest of the paper is organized in the following manner. In Section\ref{S1}
we introduce the charge operator, in Section\ref{S2} we introduce the charge
operator in presence of Gravity's Rainbow specified to the Schwarzschild and
to the de Sitter metric, in Section\ref{S3} we will apply the charge operator
to the magnetic monopole case and in Section\ref{S4} we will summarize and conclude.

\section{The Charge Operator}

\label{S1}To build the charge operator, we have to recognize the gravitational
field as a fundamental field and see what implications we have on $Q_{e}$ and
$Q_{m}$. For example, in Ref.\cite{Remo}, the r\^{o}le of $Q_{e}$ and $Q_{m}$
has been played by a cosmological constant interpreted as an eigenvalue of an
associated Sturm-Liouville problem. To do this, we have introduced the
Wheeler-DeWitt equation (WDW)\cite{De Witt} by rearranging the Einstein's
field equations, to get:%
\begin{equation}
\mathcal{H}_{\Lambda}\mathcal{=}\left(  2\kappa\right)  G_{ijkl}\pi^{ij}%
\pi^{kl}-\frac{\sqrt{g}}{2\kappa}\!{}\!\left(  \,^{3}R-2\Lambda_{c}\right)
=0,\label{HL}%
\end{equation}
for the sourceless case and in presence of a cosmological term.%
\begin{equation}
\mathcal{H}_{Q}\mathcal{=}\left(  2\kappa\right)  G_{ijkl}\pi^{ij}\pi
^{kl}-\frac{\sqrt{g}}{2\kappa}\!{}\!\left(  \,^{3}R-\mathcal{H}_{M}\right)
=0,\label{HQ}%
\end{equation}
with a matter term and in absence of a cosmological constant. Note the formal
similarity between Eqs.$\left(  \ref{HL}\right)  $ and $\left(  \ref{HQ}%
\right)  $. $G_{ijkl}$ is the \textit{supermetric} defined as%
\begin{equation}
G_{ijkl}=\frac{1}{2\sqrt{g}}(g_{ik}g_{jl}+g_{il}g_{jk}-g_{ij}g_{kl})
\end{equation}
and $^{3}R$ is the scalar curvature in three dimensions. $\pi^{ij}$ is called
the supermomentum. This is the time-time component of the Einstein's field
equations. It represents the invariance under \textit{time} reparametrization
and it works as a constraint at the classical level. Fixing our attention on
the constraint $\left(  \ref{HQ}\right)  $, the explicit form of
$\mathcal{H}_{M}$ is easily obtained with the help of Eqs.$\left(
\ref{elec}\right)  $ and $\left(  \ref{magn}\right)  $, where one finds%
\begin{equation}
\mathcal{H}_{M}=2\kappa T_{\alpha\beta}u^{\alpha}u^{\beta}=\frac{\kappa}{4\pi
}\frac{Q_{e}^{2}+Q_{m}^{2}}{r^{4}}.
\end{equation}
Thus, the classical constraint $\mathcal{H}_{Q}$ becomes%
\begin{equation}
\mathcal{H}_{Q}\mathcal{=}\left(  2\kappa\right)  G_{ijkl}\pi^{ij}\pi
^{kl}-\frac{\sqrt{g}}{2\kappa}\!{}\!\,\left(  ^{3}R-\frac{\kappa}{4\pi}%
\frac{Q_{e}^{2}+Q_{m}^{2}}{r^{4}}\right)  =0.
\end{equation}
For a spherically symmetric metric described by $\left(  \ref{dS}\right)  $
with $g_{1}\left(  E/E_{P}\right)  =g_{2}\left(  E/E_{P}\right)  =1$, it is
easy to recognize that the classical constraint reduces to%
\begin{equation}
^{3}R=2G\frac{Q_{e}^{2}+Q_{m}^{2}}{r^{4}}\qquad\Longrightarrow\qquad
b^{\prime}\left(  r\right)  =G\frac{Q_{e}^{2}+Q_{m}^{2}}{r^{2}},\label{Hcl}%
\end{equation}
whose solution represents the Reissner-Nordstr\"{o}m (RN) metric if%
\begin{equation}
N^{2}\left(  r\right)  =\left[  1-\frac{b\left(  r\right)  }{r}\right]  ^{-1}%
\end{equation}
and%
\begin{equation}
b\left(  r\right)  =2MG-\frac{G\left(  Q_{e}^{2}+Q_{m}^{2}\right)  }%
{r}.\label{p12}%
\end{equation}
On the other hand, changing the point of view, one could fix the background to
see if there are other combinations solving the classical constraint $\left(
\ref{Hcl}\right)  $. For example, if one fixes the background metric to be the
Schwarzschild metric, one finds that the only solution compatible with the
classical constraint is the trivial solution $Q_{e}=Q_{m}=0$. The same
situation happens for the de Sitter (dS) and Anti-de Sitter (AdS) metric.
Things can change if we consider quantum fluctuations of the gravitational
field. Indeed, these can be a source of nontrivial solutions as shown in Ref.
\cite{RemoCharge}. To this purpose, we promote $\mathcal{H}_{Q}$ to be an
operator and the WDW equation in the presence of an electromagnetic field
becomes
\begin{equation}
\mathcal{H}_{Q}\Psi=\left[  \left(  2\kappa\right)  G_{ijkl}\pi^{ij}\pi
^{kl}-\frac{\sqrt{g}}{2\kappa}\!{}\!\left(  ^{3}R-\frac{\kappa}{4\pi}%
\frac{Q_{e}^{2}+Q_{m}^{2}}{r^{4}}\right)  \right]  \Psi=0.\label{WDW}%
\end{equation}
The WDW equation can be cast into the form
\begin{equation}
\hat{Q}_{\Sigma}\Psi\lbrack g_{ij}]=-\frac{\sqrt{g}}{8\pi r^{4}}\left(
Q_{e}^{2}+Q_{m}^{2}\right)  \Psi\lbrack g_{ij}],\label{wdw1}%
\end{equation}
where%
\begin{equation}
\hat{Q}_{\Sigma}=\left(  2\kappa\right)  G_{ijkl}\pi^{ij}\pi^{kl}-\frac
{\sqrt{g}}{2\kappa}\!{}\,^{3}R
\end{equation}
is the charge operator. Now we see that this equation formally looks like an
eigenvalue equation. To further proceed, we multiply Eq.$\left(
\ref{wdw1}\right)  $ by $\Psi^{\ast}\left[  g_{ij}\right]  $ and we
functionally integrate over the three spatial metric $g_{ij}$, to obtain
\begin{equation}
\int\mathcal{D}\left[  g_{ij}\right]  \Psi^{\ast}\left[  g_{ij}\right]
\hat{Q}_{\Sigma}\Psi\left[  g_{ij}\right]  =-\frac{1}{8\pi}\int\mathcal{D}%
\left[  g_{ij}\right]  \Psi^{\ast}\left[  g_{ij}\right]  \left(  \,\sqrt
{g}\frac{Q_{e}^{2}+Q_{m}^{2}}{r^{4}}\right)  \Psi\left[  g_{ij}\right]  .
\end{equation}
Finally one can formally re-write the WDW equation as%
\begin{equation}
\frac{\left\langle \Psi\left\vert \int_{\Sigma}d^{3}x\hat{Q}_{\Sigma
}\right\vert \Psi\right\rangle }{\left\langle \Psi|\Psi\right\rangle }%
=-\frac{Q_{e}^{2}+Q_{m}^{2}}{8\pi}\frac{\left\langle \Psi\left\vert
\int_{\Sigma}d^{3}x\left(  \,\sqrt{g}/r^{4}\right)  \right\vert \Psi
\right\rangle }{\left\langle \Psi|\Psi\right\rangle },\label{WDW0}%
\end{equation}
where we have integrated over the hypersurface $\Sigma$. The l.h.s. of
Eq.$\left(  \ref{WDW0}\right)  $ can be interpreted as an expectation value
and the r.h.s. can be regarded as the associated eigenvalue with a weight. In
principle, one should expand in perturbations even the determinant to one
loop. This means that
\begin{equation}
\frac{\left\langle \Psi\left\vert \int_{\Sigma}d^{3}x\hat{Q}_{\Sigma
}\right\vert \Psi\right\rangle }{\left\langle \Psi|\Psi\right\rangle }%
=-\frac{Q_{e}^{2}+Q_{m}^{2}}{8\pi}\frac{\left\langle \Psi\left\vert
\int_{\Sigma}d^{3}x\frac{\,\sqrt{g}^{\left(  0\right)  }+\sqrt{g}^{\left(
1\right)  }+\sqrt{g}^{\left(  2\right)  }+\ldots\sqrt{g}^{\left(  n\right)  }%
}{r^{4}}\right\vert \Psi\right\rangle }{\left\langle \Psi|\Psi\right\rangle },
\end{equation}
where $\sqrt{g}^{\left(  n\right)  }$ is the order of the approximation.
However one may also adopt an alternative approach, where one fixes the
background on the l.h.s. of Eq. $\left(  \ref{WDW0}\right)  $ and consequently
let the quantum fluctuations evolve, and then one verifies what kind of
solutions one can extract from the r.h.s. in a recursive way. Therefore the
first step begins with the r.h.s. of Eq. $\left(  \ref{WDW0}\right)  $ which
further can be reduced to
\begin{equation}
\frac{\left\langle \Psi\left\vert \int_{\Sigma}d^{3}x\hat{Q}_{\Sigma
}\right\vert \Psi\right\rangle }{\left\langle \Psi|\Psi\right\rangle }%
=-\frac{Q_{e}^{2}+Q_{m}^{2}}{8\pi}\int_{\Sigma}d^{3}x\frac{\sqrt{g}^{\left(
0\right)  }}{r^{4}}.\label{Hel}%
\end{equation}
If Eq.$\left(  \ref{Hel}\right)  $ gives the desired nontrivial eigenvalues at
zero order, it means that an electric or magnetic charge (or both) has been
created. This means that after the charge creation, the correct background
will be represented by a Reissner-Nordstr\"{o}m metric. This also means that
Eq.$\left(  \ref{Hel}\right)  $ cannot be used anymore. In the next section we
will discuss some subtleties arising in dealing with quantum fluctuations of
the determinant of the r.h.s. of Eq. $\left(  \ref{WDW0}\right)  $. If we
consider $Q_{e}^{2}\left(  Q_{m}^{2}\right)  $ as eigenvalues of the
Sturm-Liouville problem for some fixed background, we unavoidably find that
the one loop calculation is plagued by UV divergences. Therefore a
regularization/renormalization scheme is needed to remove the divergences
\cite{RemoCharge}. Nevertheless, the purpose of this paper is to propose a
procedure to avoid such a scheme: the computation of $Q_{e}^{2}\left(
Q_{m}^{2}\right)  $ in presence of Gravity's Rainbow which introduces only one
scale, the Planck scale $E_{P}$.

\section{The Charge Operator in presence of Gravity's Rainbow}

\label{S2}To compute the electric/magnetic charge in Gravity's Rainbow, we
begin with the line element $\left(  \ref{dS}\right)  $. The form of the
background is such that the \textit{shift function}
\begin{equation}
N^{i}=-Nu^{i}=g_{0}^{4i}=0
\end{equation}
vanishes, while $N$ is the previously defined \textit{lapse function}. Thus
the definition of $K_{ij}$ implies
\begin{equation}
K_{ij}=-\frac{\dot{g}_{ij}}{2N}=\frac{g_{1}\left(  E\right)  }{g_{2}%
^{2}\left(  E\right)  }\tilde{K}_{ij},
\end{equation}
where the dot denotes differentiation with respect to the time $t$ and the
tilde indicates the quantity computed in absence of Rainbow's functions
$g_{1}\left(  E\right)  $ and $g_{2}\left(  E\right)  $. For simplicity, we
have set $E_{P}=1$ in $g_{1}\left(  E/E_{P}\right)  $ and $g_{2}\left(
E/E_{P}\right)  $ but later we will bring it back for relative comparison. The
trace of the extrinsic curvature, therefore becomes%
\begin{equation}
K=g^{ij}K_{ij}=g_{1}\left(  E\right)  \tilde{K}%
\end{equation}
and the momentum $\pi^{ij}$ conjugate to the three-metric $g_{ij}$ of $\Sigma$
is%
\begin{equation}
\pi^{ij}=\frac{\sqrt{g}}{2\kappa}\left(  Kg^{ij}-K^{ij}\right)  =\frac
{g_{1}\left(  E\right)  }{g_{2}\left(  E\right)  }\tilde{\pi}^{ij}.
\end{equation}
Recalling that $u_{\mu}=\left(  -N,0,0,0\right)  $, in presence of Gravity's
Rainbow we have the following modification%
\begin{equation}
u_{\mu}=\frac{\tilde{u}_{\mu}}{g_{1}\left(  E\right)  }\qquad\Longrightarrow
\qquad u^{\mu}=g_{1}\left(  E\right)  \tilde{u}^{\mu}%
\end{equation}
which is useful to compute the distorted electromagnetic energy-momentum
tensor. Indeed, from Eqs.$\left(  \ref{elec}\right)  $ and $\left(
\ref{magn}\right)  $, we find%
\begin{equation}
T_{\mu\nu}u^{\mu}u^{\nu}=\frac{g_{2}^{2}\left(  E\right)  \tilde{g}^{11}}%
{8\pi}\left(  \tilde{F}_{01}\right)  ^{2}\tilde{u}^{\mu}\tilde{u}^{\nu}%
g_{1}^{2}\left(  E\right)  =\frac{1}{8\pi}\frac{Q_{e}^{2}}{r^{4}}g_{1}%
^{2}\left(  E\right)  g_{2}^{2}\left(  E\right)  ,
\end{equation}
for the electric charge, while when we consider the magnetic charge, we get%
\begin{equation}
T_{\mu\nu}u^{\mu}u^{\nu}=\frac{\tilde{g}_{00}}{8\pi g_{1}^{2}\left(  E\right)
}\left(  \tilde{F}_{23}\right)  ^{2}\tilde{u}^{\mu}\tilde{u}^{\nu}\tilde
{g}^{22}\tilde{g}^{33}g_{1}^{2}\left(  E\right)  g_{2}^{4}\left(  E\right)
=\frac{1}{8\pi}\frac{Q_{m}^{2}}{r^{4}}g_{2}^{4}\left(  E\right)  .
\end{equation}
Since the scalar curvature $R$ has the following property%
\begin{equation}
R=g^{ij}R_{ij}=g_{2}^{2}\left(  E\right)  \tilde{R},\label{RR}%
\end{equation}
we find that the WDW equation becomes%
\begin{equation}
\mathcal{H}\Psi=\left[  \left(  2\kappa\right)  \frac{g_{1}^{2}\left(
E\right)  }{g_{2}^{3}\left(  E\right)  }\tilde{G}_{ijkl}\tilde{\pi}^{ij}%
\tilde{\pi}^{kl}\mathcal{-}\frac{\sqrt{\tilde{g}}}{2\kappa g_{2}\left(
E\right)  }\left(  \tilde{R}-\frac{\kappa}{4\pi r^{4}}Q_{eg_{1};mg_{2}}%
^{2}\right)  \right]  \Psi=0,\label{AccaR}%
\end{equation}
where we have defined%
\begin{equation}
Q_{eg_{1};mg_{2}}^{2}=Q_{e}^{2}g_{1}^{2}\left(  E\right)  +Q_{m}^{2}g_{2}%
^{2}\left(  E\right)  \label{eg1mg2}%
\end{equation}
and where
\begin{equation}
G_{ijkl}=\frac{1}{2\sqrt{g}}\left(  g_{ik}g_{jl}+g_{il}g_{jk}-g_{ij}%
g_{kl}\right)  =\frac{\tilde{G}_{ijkl}}{g_{2}\left(  E\right)  }.
\end{equation}
By repeating the same steps that have led to Eq. $\left(  \ref{WDW0}\right)
$, we find
\begin{equation}
\frac{\langle\Psi|\int_{\Sigma}d^{3}x~\hat{Q}_{\Sigma}|\Psi\rangle}%
{\langle\Psi|\Psi\rangle}=-\frac{1}{8\pi}\frac{g_{1}^{2}(E)}{g_{2}(E)}%
\frac{\left\langle \Psi\left\vert \int_{\Sigma}d^{3}x\left(  Q_{eg_{1};mg_{2}%
}^{2}\,\sqrt{\tilde{g}}/r^{4}\right)  \right\vert \Psi\right\rangle
}{\left\langle \Psi|\Psi\right\rangle },\label{eq1}%
\end{equation}
where we have defined the distorted charge operator%
\begin{equation}
\hat{Q}_{\Sigma}=2\kappa\frac{g_{1}^{2}\left(  E\right)  }{g_{2}^{3}\left(
E\right)  }\tilde{G}_{ijkl}~\tilde{\pi}^{ij}~\tilde{\pi}^{kl}\mathcal{-}%
\frac{\sqrt{\tilde{g}}}{2\kappa g_{2}\left(  E\right)  }~\tilde{R}.
\end{equation}
Since Eq. $\left(  \ref{eq1}\right)  $ as well as Eq. $\left(  \ref{WDW0}%
\right)  $ cannot be solved exactly, we adopt a variational procedure with
trial wave functionals of the Gaussian type. To further proceed, we fix a
background metric $\bar{g}_{ij}$ and we consider quantum fluctuation around
the background of the form $g_{ij}=\bar{g}_{ij}+h_{ij}$. Following the
procedure in Ref.\cite{GaMa}, we canonically separate the degrees of freedom
and since only the transverse traceless (TT) tensor contribution becomes
relevant, we find
\begin{equation}
\hat{Q}_{\Sigma}=\frac{1}{4}\int_{\Sigma}d^{3}x\sqrt{\widetilde{\bar{g}}%
}~\tilde{G}^{ijkl}\left[  2\kappa\frac{g_{1}^{2}(E)}{g_{2}^{3}(E)}\tilde
{K}^{-1\perp}(x,x)_{ijkl}+\frac{1}{2\kappa g_{2}(E)}\{\tilde{\Delta}_{L}%
^{m}\tilde{K}^{\perp}(x,x)\}_{ijkl}\right]  ,\label{eqn2}%
\end{equation}
where we have functionally integrated over Gaussian trial wave functionals.
$\tilde{\Delta}_{L}^{m}$ represents the modified Lichnerowicz operator whose
expression is
\begin{equation}
(\hat{\Delta}_{L}^{m}h^{\perp})_{ij}=(\Delta_{L}h^{\perp})_{ij}-4~R_{i}%
^{k}~h_{kj}^{\perp}+^{3}R~h_{ij}^{\perp}~~.
\end{equation}
Now when we consider the eigenvalue equation
\begin{equation}
(\hat{\Delta}_{L}^{m}~h^{\perp})_{ij}=E^{2}~h_{ij}^{\perp}%
\end{equation}
we find
\begin{equation}
(\tilde{\Delta}_{L}^{m}\tilde{h}^{\perp})_{ij}=\frac{E^{2}}{g_{2}^{2}%
(E)}~\tilde{h}_{ij}^{\perp}~~\label{Eigen}%
\end{equation}
and the propagator $K^{\perp}(x,y)_{iakl}$ can be represented as
\begin{equation}
K^{\perp}(\vec{x},\vec{y})_{iakl}=\tilde{K}^{\perp}(\vec{x},\vec{y}%
)_{iakl}=\sum_{\tau}\frac{\tilde{h}_{ia}^{(\tau)\perp}(\vec{x})~\tilde{h}%
_{kl}^{(\tau)\perp}(\vec{y})}{2\lambda(\tau)~g_{2}^{4}(E)}%
\end{equation}
where $\tilde{h}_{ia}^{(\tau)\perp}(\vec{x})$ are the eigenfunctions of
$\tilde{\Delta}_{L}^{m}$. $\tau$ denotes a complete set of indices and
$\lambda(\tau)$ are a set of variational parameters to be determined by the
minimization of Eq.$\left(  {\ref{eqn2}}\right)  $. The expectation value of
$\hat{Q}_{\Sigma}^{\perp}$ is obtained by plugging the propagator in
Eq.$\left(  \ref{eqn2}\right)  $ and minimizing with respect to the
variational function $\lambda(\tau)$. Therefore the one-loop charge in
Gravity's Rainbow for the TT tensors is%
\begin{equation}
Q_{\Sigma}=-\frac{1}{2}\sum_{\tau}g_{1}(E)g_{2}(E)\left[  \sqrt{E_{1}^{2}%
(\tau)}+\sqrt{E_{2}^{2}(\tau)}\right]  ,\label{eqn3}%
\end{equation}
where%
\begin{equation}
Q_{\Sigma}=\frac{1}{8\pi}\frac{g_{1}^{2}(E)}{g_{2}(E)}\int_{\Sigma}%
d^{3}x~\sqrt{^{3}\tilde{g}}~\frac{Q_{eg_{1};mg_{2}}^{2}}{r^{4}}.
\end{equation}
It is important to remark that if we had considered quantum fluctuations of
the r.h.s. of Eq. $\left(  \ref{WDW0}\right)  $, then the r.h.s. of Eq.
$\left(  \ref{Eigen}\right)  $ would have been modified with the introduction
of the charge term which is possible only when Eq. $\left(  \ref{Hel}\right)
$ is solved. The expression in Eq.$\left(  \ref{eqn3}\right)  $ makes sense
when $E_{i}^{2}(\tau)>0$, where $E_{i}^{2}$ are the eigenvalues of
$\tilde{\Delta}_{L}^{m}$. Using the WKB approximation as used by 't Hooft in
the brick wall problem we can evaluate Eq.$\left(  \ref{eqn3}\right)  $
explicitly. Extracting the energy density, we can write
\begin{equation}
\frac{1}{2}\frac{g_{1}^{2}(E)}{g_{2}(E)}\frac{Q_{eg_{1};mg_{2}}^{2}}{r^{4}%
}=-\frac{1}{3\pi^{2}}\sum_{i=1}^{2}\int_{E^{\ast}}^{\infty}E_{i}g_{1}%
(E)g_{2}(E)\frac{d}{dE_{i}}\left[  \frac{E_{i}^{2}}{g_{2}^{2}(E)}-m_{i}%
^{2}(r)\right]  ^{\frac{3}{2}}dE_{i},\label{eqn4}%
\end{equation}
where $E^{\ast}$ is the value which annihilates the argument of the root and
where we have defined two r-dependent effective masses $m_{1}^{2}\left(
r\right)  $ and $m_{2}^{2}\left(  r\right)  $%
\begin{equation}
\left\{
\begin{array}
[c]{c}%
m_{1}^{2}\left(  r\right)  =\frac{6}{r^{2}}\left(  1-\frac{b\left(  r\right)
}{r}\right)  +\frac{3}{2r^{2}}b^{\prime}\left(  r\right)  -\frac{3}{2r^{3}%
}b\left(  r\right)  \\
\\
m_{2}^{2}\left(  r\right)  =\frac{6}{r^{2}}\left(  1-\frac{b\left(  r\right)
}{r}\right)  +\frac{1}{2r^{2}}b^{\prime}\left(  r\right)  +\frac{3}{2r^{3}%
}b\left(  r\right)
\end{array}
\right.  \quad\left(  r\equiv r\left(  x\right)  \right)  .\label{masses}%
\end{equation}
We have hitherto used a generic form of the background. We now fix the
attention on some backgrounds which have the following property%
\begin{equation}
m_{0}^{2}\left(  r\right)  =m_{2}^{2}\left(  r\right)  =-m_{1}^{2}\left(
r\right)  ,\qquad\forall r\in\left(  r_{t},r_{1}\right)  .\label{eqn5}%
\end{equation}
For example, the Schwarzschild background represented by the choice $b\left(
r\right)  =r_{t}=2MG$ satisfies the property $\left(  \ref{eqn5}\right)  $ in
the range $r\in\left[  r_{t},5r_{t}/4\right]  $. Similar backgrounds are the
Schwarzschild-de Sitter and Schwarzschild-Anti-de Sitter. On the other hand,
other backgrounds, like dS, AdS and Minkowski have the property%
\begin{equation}
m_{0}^{2}\left(  r\right)  =m_{2}^{2}\left(  r\right)  =m_{1}^{2}\left(
r\right)  ,\qquad\forall r\in\left(  r_{t},\infty\right)  .\label{equal}%
\end{equation}

\subsection{The Schwarzschild Case}

\label{S2S}Before going on, we examine the classical constraint for the
Schwarzschild metric. From Eq.$\left(  \ref{AccaR}\right)  $, the condition
$\mathcal{H}=0$ reduces to%
\begin{equation}
\tilde{R}-\kappa\frac{Q_{eg_{1};mg_{2}}^{2}}{4\pi r^{4}}=0\qquad
\Longrightarrow\qquad\frac{Q_{e}^{2}}{r^{4}}g_{1}^{2}\left(  E\right)
+\frac{Q_{m}^{2}}{r^{4}}g_{2}^{2}\left(  E\right)  =0,
\end{equation}
leading to the only trivial classical solution $Q_{e}=Q_{m}=0$. Note that even
in Minkowski space, we have a trivial solution with vanishing charges. This
situation persists even at the quantum level, because there is no parameter
which fixes the scale like the Schwarzschild mass can do. To further proceed,
we observe that the Schwarzschild background satisfies condition $\left(
\ref{eqn5}\right)  $ and Eq. $\left(  \ref{eqn4}\right)  $ becomes
\begin{equation}
\frac{1}{2}\frac{g_{1}^{2}(E/E_{P})}{g_{2}(E/E_{P})}\frac{Q_{eg_{1};mg_{2}%
}^{2}}{r^{4}}=-\frac{1}{3\pi^{2}}\left(  I_{+}+I_{-}\right)  ,\label{eqn51}%
\end{equation}
where%
\begin{equation}
I_{+}=3\int_{0}^{\infty}E^{2}g_{1}\left(  E/E_{P}\right)  \sqrt{\frac{E^{2}%
}{g_{2}^{2}\left(  E/E_{P}\right)  }+m_{0}^{2}\left(  r\right)  }\frac{d}%
{dE}\left(  \frac{E}{g_{2}\left(  E/E_{P}\right)  }\right)  dE\label{I+}%
\end{equation}
and%
\begin{equation}
I_{-}=3\int_{E^{\ast}}^{\infty}E^{2}g_{1}\left(  E/E_{P}\right)  \sqrt
{\frac{E^{2}}{g_{2}^{2}\left(  E/E_{P}\right)  }-m_{0}^{2}\left(  r\right)
}\frac{d}{dE}\left(  \frac{E}{g_{2}\left(  E/E_{P}\right)  }\right)
dE.\label{I-}%
\end{equation}
For convenience we have reintroduced the Planck energy scale $E_{p}$ in Eqs.
$\left(  \ref{I+}\right)  $ and $\left(  \ref{I-}\right)  $. It is clear that
the final result is strongly dependent on the choices we can do about
$g_{1}\left(  E/E_{P}\right)  $ and $g_{2}\left(  E/E_{P}\right)  $.
Nevertheless, some classes of functions cannot be considered because they do
not lead to a finite result. For example, fixing
\begin{equation}
g_{1}\left(  E/E_{P}\right)  =1-\eta\left(  E/E_{P}\right)  ^{n}%
\qquad\text{and}\qquad g_{2}\left(  E/E_{P}\right)  =1,
\end{equation}
with $\eta$ a dimensionless parameter and $n$ an integer \cite{g1g2}, Eq.
$\left(  \ref{eqn51}\right)  $ does not lead to a finite result and therefore
will be discarded. Other examples that we have to discard without involving a
specific form of $g_{1}\left(  E/E_{P}\right)  $ and $g_{2}\left(
E/E_{P}\right)  $ are:
\begin{equation}
g_{2}(E/E_{P})=g_{1}^{4}(E/E_{P}),\label{Choice a}%
\end{equation}%
\begin{equation}
g_{2}(E/E_{P})=g_{1}^{-2}(E/E_{P})\label{Choice b}%
\end{equation}
and
\begin{equation}
g_{2}^{-2}(E/E_{P})=g_{1}(E/E_{P}).\label{Choice c}%
\end{equation}
When we adopt the choice $\left(  \ref{Choice a}\right)  $, the electric
charge becomes independent on the Rainbow's functions and Eq. $\left(
\ref{eqn51}\right)  $ becomes
\begin{equation}
\frac{1}{2r^{4}}\left(  Q_{e}^{2}+Q_{m}^{2}g_{1}^{6}\left(  E/E_{P}\right)
\right)  =-\frac{1}{3\pi^{2}}\left(  I_{+}+I_{-}\right)  .
\end{equation}
In order to have real results, the argument of the square root in $I_{-}$ must
be positive for $E\gg E_{P}$ and this happens when $g_{1}(E/E_{P})$ is of the
form
\begin{equation}
g_{1}(E/E_{P})=\sqrt[4]{1+E/E_{P}}%
\end{equation}
leading to a divergent result. The same situation happens for choice $\left(
\ref{Choice b}\right)  $ where the magnetic charge becomes independent on the
Rainbow's functions
\begin{equation}
\frac{1}{2r^{4}}\left(  Q_{e}^{2}g_{1}^{6}\left(  E/E_{P}\right)  +Q_{m}%
^{2}\right)  =-\frac{1}{3\pi^{2}}\left(  I_{+}+I_{-}\right)  .
\end{equation}
To have real results for $E\gg E_{P}$ we have to impose
\begin{equation}
g_{1}(E/E_{P})=\left(  1+E/E_{P}\right)  ^{-\frac{1}{2}}%
\end{equation}
but also in this case $I_{+}$ and $I_{-}$ diverge. Finally for the choice
$\left(  \ref{Choice c}\right)  $ we find that the integrals $I_{+}$ and
$I_{-}$ in Eq. $\left(  \ref{eqn51}\right)  $ become finite but with a
negative sign in front of the r.h.s. of Eq. $\left(  \ref{eqn51}\right)  $
which means that $Q_{e}^{2}$ $\left(  Q_{m}^{2}\right)  $ should be everywhere
negative, a result which is not compatible with observation. Since the choices
$\left(  \ref{Choice a},\ldots\ref{Choice c}\right)  $ do not give the desired
result, we need to fix independently the form of $g_{1}(E/E_{P})$ and
$g_{2}(E/E_{P})$. It is immediate to observe that $g_{1}(E/E_{P})$ must have a
shape such that $I_{+}$ and $I_{-}$ be convergent. Following Ref.\cite{GaMa},
we consider
\begin{equation}
g_{1}\left(  E/E_{P}\right)  =(1+\beta\frac{E}{E_{P}})\exp(-\alpha\frac{E^{2}%
}{E_{P}^{2}})\qquad g_{2}\left(  E/E_{P}\right)  =1\qquad\alpha,\beta\in%
\mathbb{R}
.\label{a)}%
\end{equation}
In this configuration, we know that the integrals $I_{+}$ and $I_{-}$ are
finite and the introduction of the parameter $\beta$ allows to change the sign
in the one loop term. Note that the choice of the Gaussian was dictated by a
comparison with a cosmological constant computation in the framework of
Noncommutative theory\cite{RGPN}. By defining%
\begin{equation}
x=\sqrt{\frac{m_{0}^{2}\left(  r\right)  }{E_{P}^{2}}}\label{x}%
\end{equation}
and following the same steps of Ref.\cite{GaMa}, one finds%
\begin{equation}
\frac{Q_{e}^{2}g_{1}^{4}(E/E_{P})+Q_{m}^{2}g_{1}^{2}(E/E_{P})}{2E_{P}^{4}%
r^{4}}=-\frac{1}{2\pi^{2}}f\left(  \alpha,\beta;x\right)  {,}\label{LG}%
\end{equation}
where%
\[
f\left(  \alpha,\beta;x\right)  =\left[  \frac{x^{2}}{\alpha}\cosh\left(
\frac{\alpha x^{2}}{2}\right)  K_{1}\left(  \frac{\alpha x^{2}}{2}\right)
\right.
\]%
\begin{equation}
\left.  +\beta\left(  {\frac{3x}{2{\alpha}^{2}}}-\frac{x^{2}\sqrt{\pi}%
}{{\alpha}^{\frac{3}{2}}}\sinh\left(  \alpha x^{2}\right)  +\frac{3\sqrt{\pi}%
}{2{\alpha}^{\frac{5}{2}}}\cosh\left(  \alpha x^{2}\right)  +\frac{\sqrt{\pi}%
}{2{\alpha}^{\frac{3}{2}}}\left(  x^{2}-\,{\frac{3}{2{\alpha}}}\right)
\,e{^{\alpha x^{2}}}\operatorname{erf}\left(  \sqrt{\alpha\,}x\right)
\right)  \right]  \label{f(x)}%
\end{equation}
and where $K_{1}\left(  x\right)  $ is the Bessel function of the first kind
and $\operatorname{erf}\left(  x\right)  $ is the error function. For the
Schwarzschild background, Eq.$\left(  \ref{x}\right)  $ becomes%
\begin{equation}
x=\sqrt{\frac{m_{0}^{2}\left(  r\right)  }{E_{P}^{2}}}=\left\{
\begin{array}
[c]{cc}%
\sqrt{\frac{3MG}{r^{3}E_{P}^{2}}} & r>2MG\\
\sqrt{\frac{3}{8\left(  MG\right)  ^{2}E_{P}^{2}}} & r=2MG
\end{array}
\right.
\end{equation}
and its behavior is%
\begin{equation}
x\rightarrow\left\{
\begin{array}
[c]{c}%
\infty\qquad\text{\textrm{when}}\qquad M\rightarrow0\qquad
\text{\textrm{for\qquad}}r=2MG\\
0\qquad\text{\textrm{when}}\qquad M\rightarrow0\qquad\text{\textrm{for\qquad}%
}r>2MG
\end{array}
\right.  ,\label{small}%
\end{equation}
while%
\begin{equation}
x\rightarrow\left\{
\begin{array}
[c]{c}%
0\qquad\text{\textrm{when}}\qquad M\rightarrow\infty\qquad
\text{\textrm{for\qquad}}r=2MG\\
\infty\qquad\text{\textrm{when}}\qquad M\rightarrow\infty\qquad
\text{\textrm{for\qquad}}r>2MG
\end{array}
\right.  .\label{large}%
\end{equation}
The behavior in Eq.$\left(  \ref{large}\right)  $ will be discarded, because
it does not represent a physical realization. Therefore, we fix our attention
on Eq.$\left(  \ref{small}\right)  $. For large $x$, the r.h.s. of Eq.$\left(
\ref{LG}\right)  $ becomes:%
\begin{equation}
\frac{g_{1}^{2}(E/E_{P})Q_{eg_{1};mg_{2}}^{2}}{2E_{P}^{4}r^{4}}\simeq
-{\frac{\left(  2\beta{\alpha}^{3/2}+\sqrt{\pi}{\alpha}^{2}\right)  x}%
{4\pi^{2}{\alpha}^{7/2}}-\frac{8\beta{\alpha}^{5/2}+3\sqrt{\pi}{\alpha}^{3}%
}{16\pi^{2}{\alpha}^{11/2}x}+\frac{3}{128\pi^{2}}}\,{\frac{16\beta{\alpha
}^{7/2}+5\sqrt{\pi}{\alpha}^{4}}{{\alpha}^{15/2}{x}^{3}}}+O\left(
x^{-4}\right)  ,\label{AsL}%
\end{equation}
while for small $x$ we obtain%
\begin{equation}
\frac{g_{1}^{2}(E/E_{P})Q_{eg_{1};mg_{2}}^{2}}{2E_{P}^{4}r^{4}}\simeq
-{\frac{4{\alpha}^{5/2}+3\sqrt{\pi}\beta{\alpha}^{2}}{4\pi^{2}{\alpha}^{9/2}%
}-\frac{\left(  2\sqrt{\alpha}\gamma+2\sqrt{\alpha}\ln\left(  \frac{x^{2}}%
{4}\alpha\sqrt{e}\right)  -2\sqrt{\pi}\beta\right)  x^{4}}{16\pi^{2}%
\sqrt{\alpha}}-}\frac{2\beta}{15\pi^{2}}{x}^{5}+O\left(  x^{7}\right)
,\label{SmL}%
\end{equation}
where $\gamma$ is the Euler's constant. To keep the procedure as general as
possible, in Eq. $\left(  \ref{LG}\right)  $ we have kept the combination
between the electric and magnetic charge coming from the energy-momentum
tensor expressed in Eq. $\left(  \ref{eg1mg2}\right)  $. It is also
interesting to note that every choice we can do on the function $g_{1}%
(E/E_{P})$ satisfying the assumption $\left(  \ref{a)}\right)  $, the magnetic
monopole is less suppressed in the trans-Planckian region with respect to the
electric charge. This could confirm that the magnetic monopole is a problem
related to the very early universe. On the other hand, when we are on the
cis-Planckian region, the electric charge and the magnetic monopole are not
suppressed by the Rainbow's functions and behave in the same way. For this
reason, we can first study the electric charge setting $Q_{m}^{2}=0$. It is
straightforward to see that if we fix
\begin{equation}
\beta=-{\frac{\sqrt{\alpha\pi}}{2}},\label{as}%
\end{equation}
then the linear divergent term of the asymptotic expansion $\left(
\ref{AsL}\right)  $ disappears and Eq. $\left(  \ref{LG}\right)  $ vanishes
for large $x$, namely when $r=r_{t}=2MG$ and $M\rightarrow0$. Therefore on the
throat $r_{t}$ one gets%
\begin{equation}
Q_{e}^{2}\left(  \alpha,\beta,r_{t}\right)  =-\frac{r_{t}^{4}E_{P}^{4}}%
{\pi^{2}}f\left(  \alpha,-{\frac{\sqrt{\alpha\pi}}{2}};\frac{3}{2r_{t}%
^{2}E_{P}^{2}}\right)  .\label{AsL1}%
\end{equation}
By imposing that,
\begin{equation}
Q_{e}^{2}\left(  \alpha,\beta,\bar{r}_{t}\right)  =\frac{1}{137}%
=\allowbreak0.7\,3\times10^{-2},\label{fst}%
\end{equation}
then we find%
\begin{equation}
\bar{r}_{t}=0.295/E_{P},
\end{equation}
where we have fixed $\alpha=1/4$. The situation is not much different if we
choose%
\begin{equation}
\beta=-{\frac{4\,}{3}}\sqrt{\frac{\alpha}{\pi}}.\label{sm}%
\end{equation}
Indeed, Eq.$\left(  \ref{LG}\right)  $ becomes%
\begin{equation}
Q_{e}^{2}\left(  \alpha,\beta,r_{t}\right)  =-\frac{r_{t}^{4}E_{P}^{4}}%
{\pi^{2}}f\left(  \alpha,-{\frac{4\,}{3}}\sqrt{\frac{\alpha}{\pi}};\frac
{3}{2r_{t}^{2}E_{P}^{2}}\right)  \label{SmL1}%
\end{equation}
and fixing again $Q_{e}^{2}$ like in Eq.$\left(  \ref{fst}\right)  $, one
finds%
\begin{equation}
\bar{r}_{t}=0.571/E_{P}.
\end{equation}
Both the solution require a sub-Planckian throat. It is clear that the
comparison of the fine structure constant $1/137$ with $Q_{m}^{2}$ is not
possible. However later we will discuss how the charge operator $Q_{\Sigma}$
can give information about the magnetic monopole. If we move to the region
where $5r_{t}/4>r>r_{t}$, we introduce a dependence on the radius $r$, which
can be eliminated with the computation of%
\begin{equation}
\frac{dQ_{e}^{2}}{dr}=0.\label{dQ}%
\end{equation}
However when we choose the parametrization $\left(  \ref{as}\right)  $ , the
solution of Eq.$\left(  \ref{dQ}\right)  $ is imaginary and therefore will be
discarded. On the other hand, when we choose parametrization $\left(
\ref{sm}\right)  $, we find that the expression $\left(  \ref{SmL}\right)  $
reduces to%
\begin{equation}
Q_{e}^{2}=-\frac{1}{4{\pi}^{2}}{\ln\left(  \frac{3MG}{4r^{3}E_{P}^{2}}\alpha
e^{\gamma+11/6}\right)  }\frac{\left(  3MG\right)  ^{2}}{r^{2}}+O\left(
\left(  2MG\right)  ^{5/2}\right)  \label{AsLSm2}%
\end{equation}
and the computation of Eq.$\left(  \ref{dQ}\right)  $ in this case leads to
the following relationship%
\begin{equation}
\bar{r}^{3}=\frac{3r_{t}\alpha}{8E_{P}^{2}}\exp\left(  \gamma+{\frac{10}{3}%
}\right)  .
\end{equation}
Since $r\in\left[  r_{t},5r_{t}/4\right]  $, this implies that%
\begin{equation}
r\in\left[  \sqrt{\frac{3\alpha}{8E_{P}^{2}}\exp\left(  \gamma+{\frac{10}{3}%
}\right)  },\frac{5}{4}\sqrt{\frac{3\alpha}{8E_{P}^{2}}\exp\left(
\gamma+{\frac{10}{3}}\right)  }\right]  ,
\end{equation}
\newline then we find the following bound%
\begin{equation}
2.\,\allowbreak963\,8\times10^{-2}=\frac{3^{\frac{8}{3}}}{64\pi^{2}}\geq
Q_{e}^{2}\left(  \alpha,\bar{r}\right)  \geq\frac{3^{\frac{8}{3}}}{64\pi^{2}%
}\sqrt[3]{\left(  \frac{5}{4}\right)  ^{4}}=\allowbreak3.\,\allowbreak
990\,8\times10^{-2}.
\end{equation}

\subsection{The de Sitter (Anti-de Sitter) Case}

Even in this case, we examine the classical constraint for the dS and AdS
metric, respectively. For the AdS metric it is immediate to verify that the
condition $\mathcal{H}=0$ reduces to%
\begin{equation}
\tilde{R}-\kappa\frac{Q_{eg_{1};mg_{2}}^{2}}{4\pi r^{4}}=0\qquad
\Longrightarrow\qquad G\left(  \frac{Q_{e}^{2}}{r^{4}}g_{1}^{2}\left(
E\right)  +\frac{Q_{m}^{2}}{r^{4}}g_{2}^{2}\left(  E\right)  \right)
=-\Lambda_{AdS},
\end{equation}
which is never satisfied, while for the dS metric, we find%
\begin{equation}
\tilde{R}-\kappa\frac{Q_{eg_{1};mg_{2}}^{2}}{4\pi r^{4}}=0\qquad
\Longrightarrow\qquad G\left(  \frac{Q_{e}^{2}}{r^{4}}g_{1}^{2}\left(
E\right)  +\frac{Q_{m}^{2}}{r^{4}}g_{2}^{2}\left(  E\right)  \right)
=\Lambda_{dS}.
\end{equation}
Moreover, if we fix the radius to the value $r=\sqrt{3/\Lambda_{dS}}$, we find%
\begin{equation}
G\Lambda_{dS}\left(  Q_{e}^{2}g_{1}^{2}\left(  E\right)  +Q_{m}^{2}g_{2}%
^{2}\left(  E\right)  \right)  =9,
\end{equation}
which fixes the values of $Q_{e}$, $Q_{m}$ and $\Lambda_{dS}$ to values
incompatible with observation. However, things can be different from the
quantum point of view. Since in the dS and AdS cases, the condition $\left(
\ref{equal}\right)  $ holds, Eq.$\left(  \ref{eqn4}\right)  $ becomes%
\begin{equation}
\frac{1}{2}\frac{g_{1}^{2}(E/E_{P})}{g_{2}(E/E_{P})}\frac{Q_{eg_{1};mg_{2}%
}^{2}}{r^{4}}=-\frac{2}{3\pi^{2}}I_{-},\label{eqn52}%
\end{equation}
where $I_{-}$ is given by Eq.$\left(  \ref{I-}\right)  $. Choosing the
Rainbow's functions like in the Schwarzschild case, one finds%
\begin{equation}
\frac{g_{1}^{2}(E/E_{P})Q_{eg_{1};mg_{2}}^{2}}{2E_{P}^{4}r^{4}}=-\frac{\beta
}{4\alpha^{\frac{5}{2}}\pi^{\frac{3}{2}}}(3+2\alpha x^{2})e^{-\alpha x^{2}%
}-\frac{x^{2}}{2\alpha\pi^{2}}e^{-\frac{\alpha x^{2}}{2}}K_{1}\left(
\frac{\alpha x^{2}}{2}\right)  ~~,\label{QLambda}%
\end{equation}
where $x$ is expressed by Eq.$\left(  \ref{x}\right)  $, but with a different
$m_{0}^{2}(r)$. Indeed, we have
\begin{equation}
x=\sqrt{\frac{m_{0}^{2}(r)}{E_{P}^{2}}=}\frac{1}{E_{P}r}\left\{
\begin{array}
[c]{ccc}%
\sqrt{6-\Lambda_{dS}r^{2}} & \text{\textrm{de Sitter}} & b\left(  r\right)
=\Lambda_{dS}r^{3}/3\\
&  & \\
\sqrt{6+\Lambda_{AdS}r^{2}} & \text{\textrm{Anti-de Sitter}} & b\left(
r\right)  =-\Lambda_{AdS}r^{3}/3
\end{array}
\right.  .
\end{equation}
We can gain more information by evaluating the r.h.s. of Eq.$\left(
\ref{QLambda}\right)  $ for small and large $x$. For large $x$, one gets%
\begin{equation}
\frac{g_{1}^{2}(E/E_{P})Q_{eg_{1};mg_{2}}^{2}}{2E_{P}^{4}r^{4}}\simeq
e^{-\alpha x^{2}}\left[  -\frac{\beta}{2\pi^{3/2}\alpha^{3/2}}x^{2}-\frac
{1}{2\pi^{3/2}\alpha^{3/2}}x-\frac{3\beta}{4\pi^{3/2}\alpha^{5/2}}-\frac
{3}{8\pi^{3/2}\alpha^{5/2}}\frac{1}{x}+\frac{15}{64\pi^{3/2}\alpha^{7/2}}%
\frac{1}{x^{3}}+O(x^{-5})\right]  ,\label{LLarge}%
\end{equation}
while for small $x$, we get
\begin{equation}
\frac{g_{1}^{2}(E/E_{P})Q_{eg_{1};mg_{2}}^{2}}{2E_{P}^{4}r^{4}}\simeq
-\frac{(4\sqrt{\alpha}+3\beta\sqrt{\pi})}{4\pi^{2}\alpha^{5/2}}+\frac
{(2\sqrt{\alpha}+\beta\sqrt{\pi})}{4\pi^{2}\alpha^{3/2}}x^{2}-\frac{\left[
\sqrt{\alpha}\ln\left(  \frac{x^{2}\alpha}{4}\sqrt{\alpha}\right)
+\gamma\sqrt{\alpha}-\beta\sqrt{\pi}\right]  }{8\pi^{2}\sqrt{\alpha}}%
x^{4}+O(x^{6})~.\label{LSmall}%
\end{equation}
It is interesting to note that the expression is finite for every $x$.
Beginning with the dS case, we observe that the range of the radius $r$ is
$\left[  0,\sqrt{3/\Lambda_{dS}}\right]  $ and when $r\rightarrow0$,
$x\rightarrow\infty$ which is vanishing because of behavior $\left(
\ref{LLarge}\right)  $. On the other hand, when $r\rightarrow\sqrt
{6/\Lambda_{dS}}$, $x\rightarrow0$. However $r=\sqrt{6/\Lambda_{dS}}$
corresponds to a region external to the dS horizon which is unphysical and
therefore will be discarded. Rather when%
\begin{equation}
r=\sqrt{\frac{3}{\Lambda_{dS}}}\qquad\Longrightarrow\qquad x=\frac
{\sqrt{\Lambda_{dS}}}{E_{P}}.
\end{equation}
Therefore keeping the same parametrization that allows a vanishing
contribution for small $x$, Eq.$\left(  \ref{QLambda}\right)  $ becomes%
\begin{equation}
Q_{e}^{2}\left(  \alpha,-{\frac{4\,}{3}}\sqrt{\frac{\alpha}{\pi}},\frac
{\sqrt{\Lambda_{dS}}}{E_{P}}\right)  =\frac{9E_{P}^{4}}{\Lambda_{dS}^{2}%
}\left[  \frac{2}{3\alpha^{2}\pi^{2}}(3+\frac{2\alpha\Lambda_{dS}}{E_{P}^{2}%
})\exp\left(  -\frac{\alpha\Lambda_{dS}}{E_{P}^{2}}\right)  -\frac
{\Lambda_{dS}}{\alpha\pi^{2}E_{P}^{2}}\exp\left(  -\frac{\alpha\Lambda_{dS}%
}{2E_{P}^{2}}\right)  K_{1}\left(  \alpha\frac{\Lambda_{dS}}{2E_{P}^{2}%
}\right)  \right]  ,
\end{equation}
where we have excluded the trans-Planckian region which suppresses the charge
contribution. By imposing that%
\begin{equation}
Q_{e}^{2}\left(  \frac{1}{4},-{\frac{2\,}{3\sqrt{\pi}}},\frac{\sqrt
{\bar{\Lambda}_{dS}}}{E_{P}}\right)  =\frac{1}{137},
\end{equation}
we find that%
\begin{equation}
\bar{\Lambda}_{dS}\simeq16E_{P}^{2}%
\end{equation}
and the corresponding \textquotedblleft\textit{Cosmological radius}%
\textquotedblright\ becomes%
\begin{equation}
\bar{r}_{\Lambda}^{Q_{e}}=\sqrt{\frac{3}{\bar{\Lambda}_{dS}}}=\frac
{\allowbreak0.433\,01}{E_{P}}.
\end{equation}
Concerning the AdS case, we observe that since $r\in\left[  0,+\infty\right)
$, when%
\begin{equation}
r\rightarrow+\infty\qquad\rightarrow\qquad x=\frac{\sqrt{\Lambda_{AdS}}}%
{E_{P}}.
\end{equation}
and $Q_{e}^{2}\left(  1/4,-{2\,/}\left(  3\sqrt{\pi}\right)  ,\sqrt
{\Lambda_{AdS}}/E_{P}\right)  \rightarrow\infty$ and therefore will be discarded.

\section{Magnetic Monopoles}

\label{S3}As introduced in Section\ref{S1}, our calculation applies also to
magnetic monopoles. However, since we have no experimental evidence in high
energy physics, we need to use the Dirac proposal between the magnetic
monopole and the electric charge described by the relationship $\left(
\ref{ECMM}\right)  $ to fix numbers. Therefore, it is immediate to see
that$\allowbreak$%
\begin{equation}
Q_{m}=\frac{2\pi}{Q_{e}}=73.\,\allowbreak543\qquad\Longrightarrow\qquad
Q_{m}^{2}=5408.\,\allowbreak6.
\end{equation}
Since the value of $Q_{m}^{2}$ is quite large, for the Schwarzschild metric we
can use parametrization $\left(  \ref{as}\right)  $ which keeps under control
large values of $x$, while the parametrization $\left(  \ref{sm}\right)  $
will be discarded. Setting
\begin{equation}
-\frac{r_{t}^{4}E_{P}^{4}}{\pi^{2}}f\left(  \frac{1}{4},-{\frac{\sqrt{\pi}}%
{4}};\frac{3}{2r_{t}^{2}E_{P}^{2}}\right)  =5408.\,\allowbreak6,
\end{equation}
we find%
\begin{equation}
\bar{r}_{t}^{M}=\frac{6.6}{E_{P}}.
\end{equation}
Note that when we compare $\bar{r}_{t}^{Q_{m}}$ with $\bar{r}_{t}^{Q_{e}}$, we
find%
\begin{equation}
\frac{\bar{r}_{t}^{Q_{m}}}{\bar{r}_{t}^{Q_{e}}}=\frac{6.6}{0.295}%
=22.\,\allowbreak373.
\end{equation}
On the other hand, if we use the dS metric, we find%
\begin{equation}
Q_{m}^{2}\left(  \frac{1}{4},-{\frac{2\,}{3\sqrt{\pi}}},\frac{\sqrt
{\bar{\Lambda}_{dS}}}{E_{P}}\right)  =5408.\,\allowbreak6,
\end{equation}
which implies%
\begin{equation}
\bar{\Lambda}_{dS}\simeq0.024E_{P}^{2}%
\end{equation}
and the corresponding \textquotedblleft\textit{Cosmological radius}%
\textquotedblright\ becomes%
\begin{equation}
\bar{r}_{\Lambda}^{Q_{m}}=\sqrt{\frac{3}{\bar{\Lambda}_{dS}}}=\frac
{\allowbreak11.18}{E_{P}}.
\end{equation}
Once again, when we compare $\bar{r}_{\Lambda}^{Q_{m}}$ with $\bar{r}%
_{\Lambda}^{Q_{e}}$ we find%
\begin{equation}
\frac{\bar{r}_{\Lambda}^{Q_{m}}}{\bar{r}_{\Lambda}^{Q_{e}}}=\frac
{11.18}{\allowbreak0.433\,01}=25.\,\allowbreak819.
\end{equation}

\section{Conclusions}

\label{S4}In this paper we have explored the possibility that quantum
fluctuations of the gravitational field be considered as a source for the
electric/magnetic charge. The idea is not new, because it has its origin in
the Wheeler's proposal of \textquotedblleft\textit{charge without
charge}\textquotedblright\ and \textquotedblleft\textit{mass without
mass}\textquotedblright\ arising from the \textit{spacetime foam}
picture\cite{Wheeler}. Moreover, a first approach has been proposed by one of
us in Ref.\cite{RemoCharge}. What is new in this paper is that the UV
divergences are kept under control by Gravity's Rainbow which is a distortion
of spacetime activating at the Planck's scale. This distortion avoids the
introduction of any regularization/renormalization process, like in
Noncommutative theory approaches\cite{RGPN}. Note that the \textit{Rainbow's
functions} $g_{1}(E/E_{P})$ and $g_{2}(E/E_{P})$ are constrained only by the
low energy limit $\left(  \ref{lim}\right)  $ and by the request that the one
loop integrals be UV finite\cite{GaMa,Remof(R),RGFL}. It is interesting to
note that differently from the approach of Ref.\cite{RemoCharge}, here there
is not a renormalization scale $\mu_{0}$ which is free to be fixed depending
on the problem under consideration. In this approach, $\mu_{0}=E_{P}$ since
the beginning. Moreover, as shown in section \ref{S2S}, not every choice of
$g_{1}(E/E_{P})$ and $g_{2}(E/E_{P})$ is possible, otherwise the final result
could be unphysical. The choice adopted in this paper has been borrowed by the
result obtained on the estimation of the cosmological constant made in
Ref.\cite{GaMa}. Of course we have not exhausted all the possible choices, but
if one takes seriously the method of Refs.\cite{GaMa,Remof(R)}, an agreement
also with the procedure of the present paper must be found. Indeed, in
discussing an inflationary scenario governed by Gravity's Rainbow\cite{GaSa},
it appears that a different proposal has been chosen. Nevertheless, the
functions $g_{1}(E/E_{P})$ and $g_{2}(E/E_{P})$ in Ref.\cite{GaSa}, are
present under the form of a ratio and therefore a major freedom on their
choice can be introduced. It is important to note that the electric and
magnetic charges appear as a quantum effect of the gravitational field.
Indeed, the classical contribution related to the specific geometries hitherto
examined leads to $Q_{e}=Q_{m}=0$. It is also important to remark that once
the charge has been created, the only correct metric that can be used to
discuss the solutions of Eq.$\left(  \ref{WDW0}\right)  $ is the
Reissner-Nordstr\"{o}m metric. Three basic geometries have been examined. One
of these, the AdS background leads to inconsistent solutions and therefore has
been discarded. On the other hand, the Schwarzschild and the dS background
show that the computed particle radius of the electron is of the Planckian
order. This has been obtained by fixing the value of the electric charge to
the fine structure constant that, in the units we have adopted, is coincident
with the square of the electron charge. As regards the magnetic monopole,
since no direct observation at very high energies has ever been announced, we
have used the Dirac quantization rule to obtain information about the magnetic
charge and therefore recover its own particle radius. It is interesting to
note that the ratio between the magnetic monopole radius and the electron
radius $r^{Q_{m}}/r^{Q_{e}}$ is of the same order for both the Schwarzschild
and the de Sitter background. It is also interesting to observe that the
appearance of the electric charge and the magnetic monopole as a quantum
gravitational effect in the cis-Planckian region is not affected by the
Rainbow's functions at the classical level, namely the l.h.s. of Eqs.$\left(
\ref{eqn51},\ref{eqn52}\right)  $ as it should be. However in the
trans-Planckian region an asymmetry is present between the electric charge and
the magnetic charge. Indeed, the electric charge is suppressed by a factor of
$g_{1}^{4}(E/E_{P})$, while the magnetic monopole is suppressed by a factor
$g_{1}^{2}(E/E_{P})$. I\ draw the reader's attention on the property that
$g_{1}(E/E_{P})\rightarrow0$ when $E/E_{P}\rightarrow\infty$. Therefore, from
the Gravity's Rainbow point of view, it seems that the magnetic monopole in
the trans-Planckian region can survive more compared to the electric charge,
or in other terms the quantum gravitational fluctuations begin to produce a
magnetic monopole and when the energy decreases even the electric charge
begins to be produced. Recently another result relating Gravity's Rainbow and
its influence on topology change has been obtained\cite{RGFL1}. This seems to
suggest that Gravity's Rainbow can be considered as a good tool for probing
the spacetime foam picture suggested by Wheeler.

\appendix{}

\section{The electromagnetic energy-momentum tensor in SI Units}

\label{app1}The electromagnetic energy-momentum tensor in free space and in SI
Units is defined as%
\begin{equation}
T_{\mu\nu}=\frac{1}{\mu_{0}}\left[  F_{\mu\gamma}F_{\nu}^{\gamma}-\frac{1}%
{4}g_{\mu\nu}F_{\gamma\delta}F^{\gamma\delta}\right]  ,
\end{equation}
where $\mu_{0}$ is the vacuum permeability, $F_{\mu\nu}=\partial_{\mu}A_{\nu
}-\partial_{\nu}A_{\mu}$ is the electromagnetic field strength tensor. For a
pure electric field, the electromagnetic potential $A_{\mu}$ assumes the form%
\begin{equation}
A_{\mu}=\left(  \frac{Q_{e}}{4\pi\epsilon_{0}rc},0,0,0\right)  \qquad
\Longrightarrow\qquad F_{\mu\nu}=F_{01}=-\frac{Q_{e}}{4\pi\epsilon_{0}r^{2}c},
\end{equation}
where $\epsilon_{0}$ is the vacuum permittivity, $c$ is the speed of light and
$Q_{e}$ is the electric charge. On the other hand, for the pure magnetic
field, the form is%
\begin{equation}
A_{\mu}=\left(  0,0,0,-\mu_{0}\frac{Q_{m}}{4\pi}\cos\theta\right)
\qquad\Longrightarrow\qquad F_{\mu\nu}=F_{23}=\mu_{0}\frac{Q_{m}}{4\pi}%
\sin\theta,
\end{equation}
where $Q_{m}$ is the magnetic charge measured in Amp\`{e}re$\cdot$meter
(A$\cdot$m). Thus the $T_{00}$ component of the energy-momentum tensor for the
electromagnetic charges becomes%
\[
T_{00}=\frac{1}{\mu_{0}}\left\{  \frac{1}{2}g^{11}\left(  F_{01}\right)
^{2}-\frac{1}{2}g_{00}\left(  g^{22}g^{33}\right)  \left(  F_{23}\right)
^{2}\right\}  =\frac{1}{2\mu_{0}\left(  4\pi r^{2}\right)  ^{2}}\left\{
g^{11}\frac{Q_{e}^{2}}{\epsilon_{0}^{2}c^{2}}-g_{00}\mu_{0}^{2}Q_{m}%
^{2}\right\}
\]%
\begin{equation}
=\frac{1}{2\left(  4\pi r^{2}\right)  ^{2}}\left\{  g^{11}\frac{Q_{e}^{2}%
}{\epsilon_{0}}-g_{00}\mu_{0}Q_{m}^{2}\right\}  ,
\end{equation}
where we have used the following relationship $c^{2}\epsilon_{0}\mu_{0}=1$.
With the help of the time-like vector $u^{\mu}$, we obtain%
\begin{equation}
T_{\mu\nu}u^{\mu}u^{\nu}=\frac{1}{2\left(  4\pi r^{2}\right)  ^{2}}\left\{
\frac{Q_{e}^{2}}{\epsilon_{0}}+\mu_{0}Q_{m}^{2}\right\}  .
\end{equation}
For a spherically symmetric metric described by $\left(  \ref{dS}\right)  $
with $g_{1}\left(  E/E_{P}\right)  =g_{2}\left(  E/E_{P}\right)  =1$, it is
easy to recognize that the classical constraint $\left(  \ref{HQ}\right)  $
reduces to%
\begin{equation}
^{3}R=\frac{2\kappa}{c^{4}}T_{\alpha\beta}u^{\alpha}u^{\beta}\qquad
\Longrightarrow\qquad b^{\prime}\left(  r\right)  =\frac{G}{4\pi r^{2}c^{4}%
}\left\{  \frac{Q_{e}^{2}}{\epsilon_{0}}+\mu_{0}Q_{m}^{2}\right\}  ,
\end{equation}
whose solution represents the Reissner-Nordstr\"{o}m metric if%
\begin{equation}
N^{2}\left(  r\right)  =\left[  1-\frac{b\left(  r\right)  }{r}\right]  ^{-1}%
\end{equation}
and%
\begin{equation}
b\left(  r\right)  =\frac{2MG}{c^{2}}-\frac{G}{4\pi rc^{4}}\left\{
\frac{Q_{e}^{2}}{\epsilon_{0}}+\mu_{0}Q_{m}^{2}\right\}  .
\end{equation}
Note that in CGS units, one defines $\epsilon_{0}=\left(  4\pi\right)  ^{-1}$
and $\mu_{0}=4\pi$ and the energy-momentum tensor is in agreement with the
expression in $\left(  \ref{EM}\right)  $.


\begin{thebibliography}{99}                                                                                               %


\bibitem {Sakharov}A.D. Sakharov, \textsl{Dokl. Akad. Nauk Ser. Fiz.}
\textbf{177}, 70 (1967); \textsl{Sov. Phys. Doklady,} \textbf{12}, 1040
(1968); \textsl{Sov. Phys. Usp.} \textbf{34}, 394 (1991); \textsl{Gen. Rel.
Grav.} \textbf{32}, 365 (2000).

\bibitem {Wheeler}J. A. Wheeler, \textsl{Phys. Rev. }\textbf{97}, 511 (1955).

\bibitem {GAC}G. Amelino-Camelia, \textsl{Int.J.Mod.Phys. }\textbf{D 11}, 35
(2002); arXiv: gr-qc/0012051.G. Amelino-Camelia, \textsl{Phys.Lett. }\textbf{B
510}, 255 (2001). arXiv: hep-th/0012238.

\bibitem {MagSmo}J. Magueijo and L. Smolin, \textsl{Class. Quant. Grav.}
\textbf{21}, 1725 (2004). arXiv:gr-qc/0305055.

\bibitem {GaMa}R.~Garattini and G.~Mandanici, \textsl{Phys. Rev.} \textbf{D
83}, 084021 (2011); arXiv:1102.3803 [gr-qc]. R.~Garattini and G.~Mandanici,
\textsl{Phys. Rev.} \textbf{D 85}, 023507 (2012); arXiv:1109.6563 [gr-qc].

\bibitem {Remof(R)}R. Garattini,\textsl{ JCAP} \textbf{017}, 1306 (2013);
arXiv:1210.7760 [gr-qc].

\bibitem {RemoCharge}R. Garattini, \textsl{Phys. Lett.} \textbf{B 666}, 189
(2008)\textit{;} arXiv:0807.0082.

\bibitem {g1g2}Y. Ling, \textsl{JCAP} \textbf{17}, 708 (2007), gr-qc/0609129;
Y. Ling, X. Li and H. Zhang, \textsl{Mod.Phys.Lett. }\textbf{A 22}, 2749
(2007), gr-qc/0512084; Y. Ling, B. Hu and X. Li, \textsl{Phys. Rev.} \textbf{D
73}, 087702 (2006), gr-qc/0512083.

\bibitem {SpinIce}C. Castelnovo, R. Moessner and S. L. Sondhi,
\textit{Magnetic Monopoles in spin ice. }\textsl{Nature} \textbf{451}, 42-45.
10.1038/nature06433. L. D. C. Jaubert and P. C. W. Holdsworth,
\textit{Signature of magnetic monopole and Dirac string dynamics in spin ice.
}\textsl{Nature Physics} \textbf{5}, 258-261. 10.1038/NPHYS1227. S. T.
Bramwell, S.R. Giblin, S. Calder, R.Aldus, D. Prabhakaran and T. Fennell,
\textit{Measurement of the charge and current of magnetic monopoles in spin
ice. }\textsl{Nature} \textbf{461}, 956-U211. 10.1038/nature08500.

\bibitem {m1}P. A. Dirac, \textsl{Proc. Roy. Soc. Lond.} \textbf{A}
\textbf{133}, 60 (1931).

\bibitem {m2}H. Georgi and S. Glashow, \textsl{Phys. Rev. Lett.} \textbf{32},
438 (1974). G. 't Hooft, \textsl{Nucl. Phys.} \textbf{B} \textbf{79}, 276
(1974). A. M. Polyakov, \textsl{JETP Lett.} \textbf{20}, 194 (1974).

\bibitem {Remo}R. Garattini, \textsl{TSPU Vestnik} \textbf{44 N 7}, 72 (2004),
gr-qc/0409016. R. Garattini, \textsl{J. Phys. Conf. Ser. }\textbf{33}, 215
(2006), gr-qc/0510062.

\bibitem {De Witt}B. S. DeWitt, \textsl{Phys. Rev.} \textbf{160}, 1113 (1967).

\bibitem {RGPN}R. Garattini and P. Nicolini, \textsl{Phys. Rev.} \textbf{D
83}, 064021 (2011); arXiv:1006.5418 [gr-qc].

\bibitem {RemoPLB}R. Garattini, \textsl{Phys.Lett.} \textbf{B} 685 329 (2010);
arXiv:0902.3927 [gr-qc].

\bibitem {RGFL}R. Garattini and F. S. N. Lobo, \textsl{Phys. Rev.} \textbf{D
85}, 024043 (2012); arXiv:1111.5729 [gr-qc].

\bibitem {GaSa}R. Garattini and M. Sakellariadou, \textit{Does Gravity's
Rainbow induce Inflation without an Inflaton? }; arXiv:1212.4987 [gr-qc].

\bibitem {RGFL1}R. Garattini and F. S. N. Lobo, \textit{Gravity's Rainbow
induces Topology Change}; arXiv:1303.5566 [gr-qc].
\end{thebibliography}
\end{document}